\documentclass[10pt,twoside,journal,compsoc]{IEEEtran}
\usepackage{amsmath,amssymb,amsfonts}
\usepackage{algorithmic}
\usepackage{graphicx}
\usepackage{textcomp}
\usepackage{xcolor}
\usepackage{booktabs}
\usepackage{colortbl}
\usepackage{soul}
\usepackage[hyphens]{url}
\usepackage[hidelinks]{hyperref}
\usepackage{pdfpages}
\usepackage{pifont}
\usepackage[ruled,linesnumbered]{algorithm2e}

\usepackage{microtype}
\usepackage{orcidlink}

\ifCLASSOPTIONcompsoc
  \usepackage[nocompress]{cite}
\else
  \usepackage{cite}
\fi
\ifCLASSINFOpdf
\else
\fi
\newcommand{\BIBentryALTinterwordspacing}{\spaceskip=\fontdimen2\font plus
\fontdimen3\font minus \fontdimen4\font\relax}
\newcommand{\BIBentrySTDinterwordspacing}{\spaceskip=0pt\relax}

\makeatletter
\patchcmd{\@IEEEtitleabstractindextext}{\raggedright}{}{}{}
\long\def\@IEEEtitleabstractindextextbox#1{\parbox{0.922\textwidth}{#1}}
\makeatother
\begin{document}
%
\title{ML-MAWS: Alignment-Free Maximum Likelihood Phylogeny Estimation Using Minimal Absent Words}
%
%
%
%

\author{Papri~Saha, Sudipta~Kumar~Das, Anonnya~Sarkar,
        and Md.~Manzurul~Hasan
\IEEEcompsocitemizethanks{
\IEEEcompsocthanksitem The authors are with the Department of Computer Science, American International University-Bangladesh (AIUB), Dhaka 1229, Bangladesh.\protect\\
E-mail: \{25-93672-1, 25-93886-3, 25-93634-1\}@student.aiub.edu, manzurul@aiub.edu.\protect\smallskip
}%
\thanks{Manuscript received 19 June 2026.\protect\\
(Corresponding author: Md.\ Manzurul Hasan.)}}
\IEEEtitleabstractindextext{%
\begin{abstract}
Alignment-free methods in phylogenetic tree construction have major benefits in computational efficiency over alignment-based methods, but most sacrifice sequence information to pairwise distances, losing the statistical power of maximum likelihood (ML) inference. We describe ML-MAWS, an algorithm that fills this gap by encoding Minimal Absent Words (MAWs) as a binary presence/absence character matrix and estimating using an ML tree under the Lewis Mkv model using ascertainment bias correction. MAWs are obtained in linear time through the traversal of a suffix automaton. The pipeline incorporates strand-aware intersection filtering that retains only MAWs absent from both DNA orientations, entropy-based multi-length selection via Shannon entropy maximization to select the most informative lengths of MAWs, and parsimony-informative character capping to retain the most discriminative columns. We tested ML-MAWS on 14 benchmark datasets of bacterial, mitochondrial, viral, and simulated genomes with normalized Robinson-Foulds distances and  matching split distances against published reference trees.
The results show that while the binary encoding of MAWs can lead to higher topological error than continuous-valued distance baselines on closely related genomes, ML-MAWS is the first MAW-based method to provide per-branch bootstrap support and a rigorous probabilistic framework with ascertainment bias correction capabilities lacking from all existing alignment-free methods.
\end{abstract}

\begin{IEEEkeywords}
Alignment-free phylogenetics, Minimal absent words, Maximum likelihood, Binary character matrix, Phylogenetic tree reconstruction, Suffix automaton
\end{IEEEkeywords}}

\maketitle

\IEEEdisplaynontitleabstractindextext

%

\ifCLASSOPTIONcompsoc
\IEEEraisesectionheading{\section{Introduction}\label{sec:introduction}}
\else
\section{Introduction}\label{sec:introduction}
\fi
\IEEEPARstart{T}{he} reconstruction of a phylogenetic tree using genomic sequence data is a core computational biology problem that supports our intuitions regarding evolutionary connections, disease outbreak monitoring, and functional genomics \cite{felsenstein1981ml}. Conventional phylogenetic processes assume the use of multiple sequence alignment (MSA) as a precondition for tree estimation. Whereas alignment-based algorithms are highly accurate, MSA computation is time- and memory-consuming, with pairwise alignment taking an amount of time of $O(n^2)$ when the sequence length is $n$~\cite{r18}. With the increasing amount of whole-genome sequencing data, alignment-based pipelines have emerged as an important computational bottleneck \cite{r3_Zielezinski,r31}.

Alignment-free (AF) techniques have become appealing alternatives that avoid explicit sequence alignment~\cite{r31,r18,r34}. These techniques derive numerical characteristics, usually the frequency of words (or $k$-mer), directly from raw sequences and calculate pairwise dissimilarity scores to determine evolutionary relationships. AF methods are orders of magnitude faster than alignment-based methods and are naturally resilient to genome rearrangements, large-scale insertions and deletions, and subsequent lateral genetic transfer (LGT)~\cite{r12,r21}. The AFproject benchmark is a set of 74 AF methods used in 24 software tools, which shows their wide range of applications in phylogenomic inference. However, the benchmark also indicated that most AF methods are based on heuristic distance functions, which do not have a principled statistical basis, meaning that they are less accurate than alignment-based maximum likelihood (ML) or Bayesian methods ~\cite{r3_Zielezinski,r2_Morgenstern}.

Some of the common features compared in AF sequences include $k$-mer frequencies~\cite{r8_Chris-Andre,r9_Cafe,r10, r23}, spaced-word matches~\cite{r2_Morgenstern,r38,r37,r35}, common substring lengths~\cite{r4_Morgenstern,r22,r28}, and genome skimming distances~\cite{r29,r5_Klotzl}. Although these methods have demonstrated potential, they are characterized by a similar weakness: the features they compute are summarized into pairwise distances, ignoring character-level information that might be used to drive more statistically sound tree-inference algorithms.

An entirely different set of features is offered by the so-called Minimal Absent Words (MAWs). A MAW of a sequence $s$ is a word that is not a substring of $s$, but whose proper substrings are all found in $s$ ~\cite{chairungsee2012maw,crochemore2000automata}. MAWs are highly informative regarding sequence coassembly and have been demonstrated to be powerful discriminants in phylogenetic analysis. The use of MAWs to construct phylogenetic trees was initially suggested by ~\cite{chairungsee2012maw} who defined a distance measure on MAW sets in terms of their symmetric differences. Anjum et al.~\cite{r1_Anjum} later proposed CD-MAWS, which calculates the cosine distance between composition vectors based on MAW sets, and demonstrated competitive performance on the AFproject benchmark datasets~\cite{r3_Zielezinski}, as well as a low computational overhead. Ehsan et al.~\cite{ehsan2025cdmaws} further enhanced the performance of CD-MAWS by adding a suffix automaton-based MAW extraction algorithm to generate MAWs sequentially, allowing a two-pointer cosine distance calculation to avoid explicit vector construction. These advancements have enabled MAW-based phylogenetics to be realistic with datasets of several megabases of bacterial genomes.

Although effective, CD-MAWS and other MAW techniques have a critical architectural weakness: they simplify the multi-dimensional MAW information to a single pairwise distance between species pairs and proceed to use distance-based tree reconstruction (typically neighbor-joining). This pairwise aggregation is missing the character-level information that different evolutionary lineages have and cannot use the established statistical tools of maximum likelihood~\cite{felsenstein1981ml} or Bayesian inference.

Simultaneously, Zahin et al.~\cite{r19} proposed Peafowl, the first alignment-free scheme to use ML inference. Peafowl builds a binary presence/absence matrix based on $k$-mers and uses RAxML to estimate a tree~\cite{stamatakis2014raxml}. Although this is an essential conceptual step, generic $k$-mers do not have the linguistic properties of MAWs: a large fraction of $k$-mers either occur in all species (uninformative) or in just one species (phylogenetically noisy)~\cite{r32}. In addition, Peafowl fails to use ascertainment bias correction, which is necessary for analyzing matrices of only variable characters~\cite{lewis2001mk}.

In this study, we suggest a method that fills the gap between MAW-based feature extraction and maximum likelihood phylogenetic inference. The main contribution is architectural: ML-MAWS replaces the pairwise distance computation of CD-MAWS~\cite{r1_Anjum,ehsan2025cdmaws} with a complete binary character matrix formulation, enabling tree estimation under the Lewis Mkv substitution model~\cite{lewis2001mk} with ascertainment bias correction. To our knowledge, this is the first MAW-based method to provide per-branch bootstrap support within a statistical framework. The ML-MAWS has three contributions.

\textbf{--} A strand-aware intersection filter that retains only MAWs absent from both the forward and reverse-complement strands, ensuring that the character matrix reflects genuinely missing sequence patterns rather than strand-specific compositional asymmetries.

\textbf{--} We used Shannon entropy maximization to automatically determine the most informative range of MAW lengths by combining the characters of varying lengths to remove the need for manual parameter tuning.

\textbf{--} In large genomes, which generate millions of MAW characters, we rank columns in terms of informativeness by using parsimony-informative character capping and only keep the most discriminative subset, improving computational tractability and phylogenetic signal-to-noise ratio.

We tested ML-MAWS using 14 benchmark datasets (bacterial, mitochondrial, viral, and simulated genomes), 9 of which (AFproject~\cite{r3_Zielezinski}) were published reference trees. Using the normalized Robinson-Foulds distance~\cite{robinson1981rf} and matching split distance, we show that ML-MAWS achieves competitive topological accuracy on mitochondrial and viral genomes and, uniquely among alignment-free methods, provides per-branch statistical confidence that correctly reflects phylogenetic uncertainty.

The remainder of this paper is structured as follows. The ML-MAWS pipeline is outlined in Section~\ref{sec:methods}. The results of the experiment and a comparison with existing methods are provided in Section~\ref{sec:results}. The strengths and limitations of our approach are discussed in Section~\ref{sec:discussion}  and a conclusion is provided in Section~\ref{sec:conclusion}.

\section{Materials and Methods}\label{sec:methods}

Minimal Absent Words (MAWs)-based alignment-free methods have been proposed as an encouraging approach to the classical alignment-based phylogenetic inference ~\cite{chairungsee2012maw,r1_Anjum}. The CD-MAWS algorithm ~\cite{r1_Anjum} compares the pairwise cosine distances of the MAW composition vectors and recovers phylogenies with the neighbor-joining (NJ) method. Ehsan et al.~\cite{ehsan2025cdmaws}  later enhanced the computational performance of CD-MAWS by proposing a MAW extraction algorithm based on suffix automaton, which runs in linear time and a two-pointer distance calculation which takes advantage of the sorted nature of the resulting MAW sets. Although such distance-based methods are quick, they necessarily leave out character-level data when aggregating pairs, and they cannot take advantage of the statistical structure of maximum likelihood (ML) estimation~\cite{felsenstein1981ml}, which has been demonstrated to yield more accurate phylogenies than distance-based models under a large variety of evolutionary scenarios~\cite{r12}. Zahin et al.~\cite{r19} introduced Peafowl, which represents the presence/absence patterns of $k$-mers as a binary character matrix and performs ML inference using RAxML~\cite{stamatakis2014raxml}. Nevertheless, generic $k$-mers do not necessarily contain phylogenetic signals: the vast majority of $k$-mers are universal or clade-specific, providing non-informative characters that do not enhance resolution as they simply inflate the matrix~\cite{r32}.

We introduce the bridging paradigm of combining the linguistically inspired feature extraction of MAWs with the statistical rigor of maximum likelihood phylogenetic inference: we introduce \textsc{ML-MAWS}, which combines the linguistically inspired feature extraction of MAWs~\cite{chairungsee2012maw,r1_Anjum,ehsan2025cdmaws} with the statistical rigor of the maximum likelihood phylogenetic inference~\cite{felsenstein1981ml}. The point is that individual MAWs can be considered as binary characters, either present or absent in the MAW set of a particular species, and the resulting character matrix can be analyzed under the Lewis Mkv model~\cite{lewis2001mk} with ascertainment bias correction. This formulation maintains the complete information of the characters of all taxa at once, which is not the case with the pairwise distance reduction used by CD-MAWS~\cite{r1_Anjum,ehsan2025cdmaws}.

\begin{figure*}[!t]
\centering
\includegraphics[width=\textwidth]{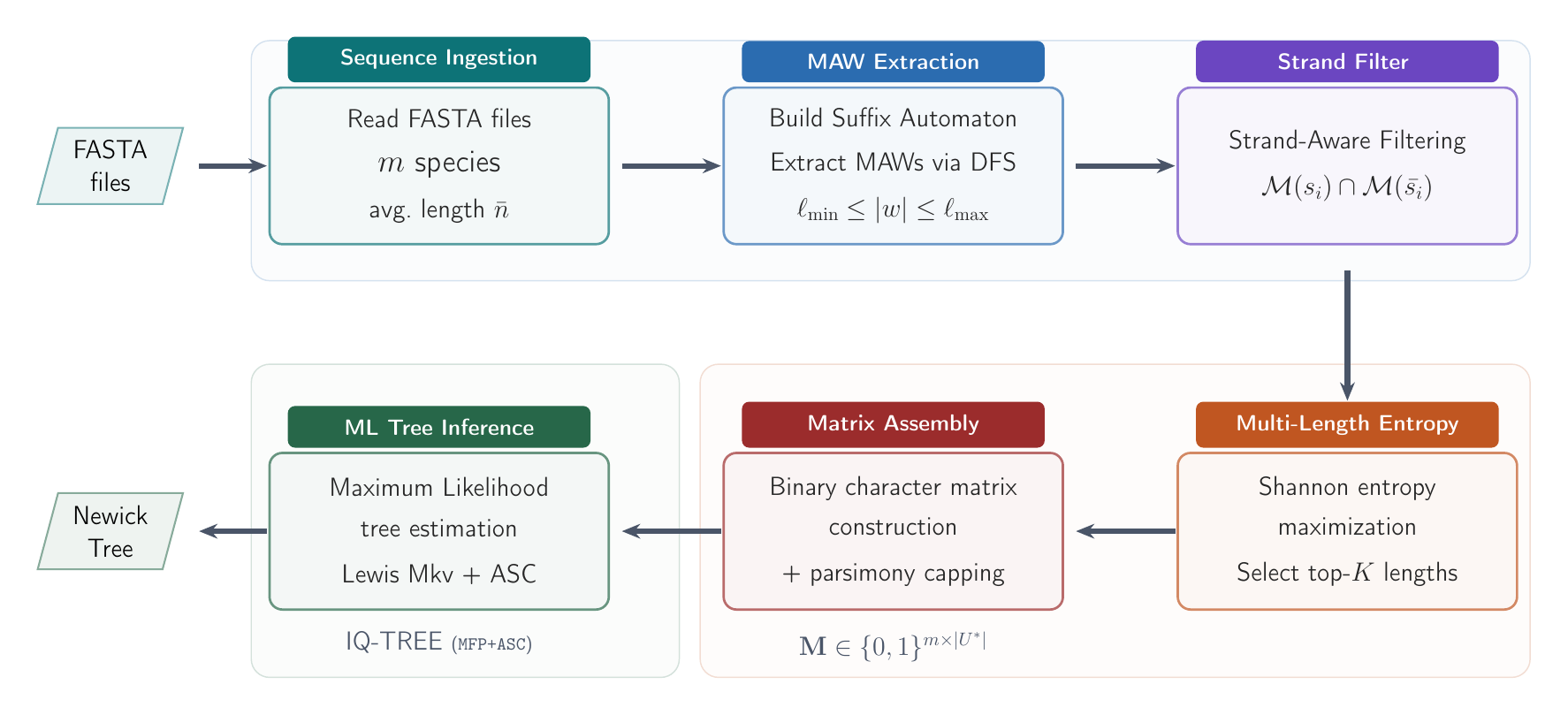}
\caption{Workflow diagram of the ML-MAWS pipeline. Genomic sequences are processed through suffix automaton--based MAW extraction with strand-aware filtering, entropy-driven multi-length selection, binary character matrix construction, and maximum likelihood tree inference via IQ-TREE.}
\label{fig:workflow_diagram}
\end{figure*}

\subsection{Notation and Preliminaries}\label{sec:notation}
Let $\sigma = \{A, C, G, T\}$ denote the DNA alphabet with $|\sigma| = 4$. For a string $s$ over $\sigma$, a word $w$ is a \emph{factor} (substring) of $s$ if $s = xwy$ for some (possibly empty) strings $x, y$. The set of all factors of $s$ is denoted $\mathrm{Fac}(s)$.
Minimal Absent Word~\cite{chairungsee2012maw}
A word $w = aub$ over $\sigma$ (where $a, b \in \sigma$) is a \emph{Minimal Absent Word} (MAW) of $s$ if $au, ub \in \mathrm{Fac}(s)$ but $w = aub \notin \mathrm{Fac}(s)$. That is, $w$ is absent from $s$ yet every proper factor of $w$ is present.

\noindent\textbf{Example.} Consider $s = \texttt{ACGT}$. The word $w = \texttt{AA}$ is a MAW of $s$ because both \texttt{A} (a proper factor of \texttt{AA}) appears in $s$, but $\texttt{AA} \notin \mathrm{Fac}(s)$.

\subsection{Sequence Ingestion}
\label{sec:stage1}
ML-MAWS takes input in either (i) a directory containing individual FASTA files, each of a species, or (ii) as a single multi-FASTA file, with each record header defining a species. In the process of parsing, ambiguity codes are eliminated (represented by $N,R,Y$) to give a clean four-letter alphabet. The average sequence length,$\bar{n}$, and species number, $m$ are measured, which will be used in the estimation of adaptive parameters later.

\subsection{MAW Extraction via Suffix Automaton}
\label{sec:stage2}

We use the MAW extraction algorithm of the suffix automaton (SA) presented by Ehsan et al.~\cite{ehsan2025cdmaws} which relies on the linear-time SA construction of Blumer et al.~\cite{blumer1985dawg}. For each species $i$, the SA for sequence $s_i$ is constructed online in $O(|s_i|)$ time and $O(|s_i|)$ space, yielding at most $2|s_i| - 1$ states and $3|s_i| - 4$ transitions~\cite{blumer1985dawg}.

\subsubsection{MAW Enumeration}
MAWs are generated by a depth-first search of the SA. Each state $v$ represents an equivalence class of right extensions of factors of $s_i$ and for each character $c \in \sigma$ with no outgoing transition ($\delta(v, c) = \varnothing$), the extension $w \cdot c$ is a candidate MAW. 
It is reported if $\ell_{\min} \leq |w \cdot c| \leq \ell_{\max}$. This process results in MAWs that are lexicographically sorted, as demonstrated by Ehsan et al.~\cite{ehsan2025cdmaws}, a feature that we can utilize later in set operations with efficient merges using two pointers.
 
\subsubsection{Strand-Aware Filtering}

Biological information is encoded in both strands of double-stranded DNA. A MAW $w$ that is absent from the forward strand $s_i$ but present as a substring of the reverse complement $\bar{s_i}$ is not truly absent from the species genome; the genomic information encoding $w$ exists on the complementary strand. Including such words as ``absent'' characters would conflate strand-specific compositional asymmetry (e.g., GC skew arising from replication bias) with genuine evolutionary signal. To address this, we construct a second suffix automaton for $\bar{s_i}$ and retain only those MAWs that are absent from both orientations:

\begin{equation}
    \mathcal{M}^{\mathrm{strand}}(s_i) = \mathcal{M}(s_i) \cap \mathcal{M}(\bar{s_i})
    \label{eq:strand}
\end{equation}

While CD-MAWS~\cite{r1_Anjum,ehsan2025cdmaws} also considers both strands during MAW extraction, it incorporates them into a pairwise cosine distance computation. ML-MAWS instead applies the intersection as an explicit filtering step on the character matrix, ensuring that each binary column represents a sequence pattern genuinely absent from the complete double-stranded genome. This minimizes the overall character count and removes strand-specific artifacts, enhancing the phylogenetic signal-to-noise ratio as shown in the ablation study (Section~\ref{sec:ablation}). To reduce the computational complexity of building two SAs per species, MAW extraction is parallelized in all $m$ species using OpenMP with dynamic scheduling.

\subsection{Entropy-Based Optimal Length Selection}
\label{sec:stage3}
The choice of the MAW length $\ell$ critically affects the phylogenetic resolution. Short MAWs ($\ell \leq 3$) tend to be universally present across all species, yielding few variable characters, whereas very long MAWs ($\ell \geq 12$) tend to be species-specific singletons. In either extreme, the phylogenetic signal is lost~\cite{chairungsee2012maw}. We addressed this using an automated, data-driven length selection procedure.

\subsubsection{Adaptive Range Estimation}

Rather than requiring the user to specify $\ell$, ML-MAWS determines a candidate range $[\ell_{\min}, \ell_{\max}]$ adaptively from the input properties. The upper bound is set as,
\begin{equation}
    \ell_{\max} = \min\!\left(\lfloor \log_2(\bar{n}) \rfloor, \; L_{\mathrm{cap}}\right)
    \label{eq:adaptive}
\end{equation}
where $\bar{n}$ is the average sequence length and $L_{\mathrm{cap}}$ is a genome-size-dependent cap (e.g., $L_{\mathrm{cap}} = 10$ for mitochondrial genomes and $L_{\mathrm{cap}} = 14$ for bacterial genomes). For datasets with $m > 50$ species, the range was expanded to increase the discriminative power.

\subsubsection{Shannon Entropy Criterion}

For each candidate length $\ell$, we construct a temporary binary matrix $\mathbf{M}_\ell$ restricted to MAWs of length exactly $\ell$ and compute the cumulative Shannon entropy over its variable columns as follows:
\begin{equation}
    H(\ell) = \sum_{j=1}^{|U_\ell|} H_j, \qquad
    H_j = -\bigl(p_j \log_2 p_j + (1{-}p_j) \log_2 (1{-}p_j)\bigr)
    \label{eq:entropy}
\end{equation}
where $p_j = \frac{1}{m}\sum_{i=1}^{m} \mathbf{M}_\ell[i][j]$ is the column frequency, and $U_\ell$ is the set of variable MAWs at length $\ell$. Constant columns ($p_j = 0$ or $p_j = 1$) contribute to zero entropy and are excluded.

\subsubsection{Multi-Length Aggregation}

Inspired by the $k$-entropy approach of Peafowl~\cite{r19}, we ranked all candidate lengths by $H(\ell)$ and selected the top-$K$ lengths (default $K = 3$), each contributing at least $\tau_{\min} = 5$ variable characters. The final character universe aggregates MAWs across the selected lengths.
\begin{equation}
    \mathcal{L}^* = \operatorname{top\text{-}K}_{\ell}\; H(\ell), \qquad
    U^* = \bigcup_{\ell \in \mathcal{L}^*} U_\ell
    \label{eq:multilength}
\end{equation}

The multi-length approach adds phylogenetically informative characters, which is especially useful for (i)small viral genomes with single-length MAWs yielding sparse matrices and (ii)highly conserved groups of species, where a single length can be insufficiently informative.

\subsection{Binary Character Matrix Construction}
\label{sec:stage4}

Within this subsection, the ML-MAWS transforms the per-species MAW sets into a unified binary character matrix with each column corresponding to a separate MAW and each row corresponding to a species. This representation is lossless, meaning that it allows for statistically principled maximum likelihood inference with branch-support quantification, as opposed to distance-based methods, which reduce pairwise relationships to a scalar.

\subsubsection{Matrix Definition}

The central innovation distinguishing ML-MAWS from distance-based approaches, such as CD-MAWS~\cite{r1_Anjum,ehsan2025cdmaws} is the construction of a binary \emph{character matrix} rather than a pairwise distance matrix. Let $U^* = \{w_1, w_2, \ldots, w_{|U^*|}\}$ denote the sorted union of all MAWs from the selected lengths across all species. The binary character matrix $\mathbf{M} \in \{0, 1\}^{m \times |U^*|}$ is defined as:
\begin{equation}
    \mathbf{M}[i][j] =
    \begin{cases}
        1 & \text{if } w_j \in \mathcal{M}^*(s_i) \\
        0 & \text{otherwise}
    \end{cases}
    \label{eq:matrix}
\end{equation}

The columns represent binary characters that encode the presence or absence of a certain MAW in each species. This expression preserves the full character-level diversification of all taxa at once and allows the likelihood-based inference of statistical interest, as opposed to the lossy aggregation of pairwise distances used by NJ-based techniques~\cite{r1_Anjum}.

\subsubsection{Efficient Construction}

Because both the per-species MAW sets and the global union $U^*$ are in sorted order, which is a property inherited from the SA-based extraction~\cite{ehsan2025cdmaws}, the matrix construction proceeds via a two-pointer merge algorithm in $O(m \cdot |U^*|)$ time. Constant columns (all-zero or all-one) are removed, as they carry no phylogenetic information under the Mkv model~\cite{lewis2001mk}.

\subsubsection{Parsimony-Informative Character Capping}

For large genomes (e.g., bacterial genomes with $|s_i| > 10^6$~bp), the character universe $|U^*|$ can exceed $10^6$, leading to prohibitive memory consumption and runtime during the ML inference. We introduce a character capping procedure: when $|U^*|$ exceeds a user-defined threshold $C_{\max}$ (default $C_{\max} = 50{,}000$), we rank columns by their parsimony informativeness, scored as $\min(n_j, m - n_j)$ where $n_j = \sum_{i=1}^{m}\mathbf{M}[i][j]$, and retain only the $C_{\max}$ highest scoring characters. 

Characters with presence frequency closest to $0.5$ carry the maximum phylogenetic signal, ensuring that the capped matrix preserves as much discriminative information as possible.
The final matrix was exported in a relaxed PHYLIP format compatible with both IQ-TREE~\cite{minh2020iqtree2} and RAxML~\cite{stamatakis2014raxml}.

\subsection{Maximum Likelihood Phylogenetic Inference}
\label{sec:stage5}
As the binary character matrix consists exclusively of variable characters (constant columns having been removed), the Lewis Mkv correction for ascertainment bias~\cite{lewis2001mk} must be applied. With IQ-TREE~\cite{minh2020iqtree2}, we invoked ModelFinder~\cite{kalyaanamoorthy2017modelfinder} using the \texttt{MFP+ASC} model specification with \texttt{-st~BIN} data type, enabling automatic selection of the best-fit binary substitution model (typically \texttt{GTR2+FO+ASC+R$k$}). With RAxML~\cite{stamatakis2014raxml}, \texttt{BINGAMMA} model can be used with the corresponding ascertainment bias correction. Branch support was assessed using $1{,}000$ replicates of the ultrafast bootstrap approximation (UFBoot2)~\cite{hoang2018ufboot2} in IQ-TREE. The resulting maximum likelihood tree was exported in the Newick format.

\subsection{Comparison with Related Methods}
\label{sec:comparison}

Table~\ref{tab:comparison} summarizes the main methodological distinctions between ML-MAWS and the two most similar alignment-free approaches.

\begin{table*}[htbp]
\centering
\caption{Feature comparison of ML-MAWS with CD-MAWS~\cite{r1_Anjum,ehsan2025cdmaws} and Peafowl~\cite{r19}. ML-MAWS combines the biologically motivated feature extraction of MAW-based methods with the statistical framework of maximum likelihood inference.}
\label{tab:comparison}
\begin{tabular}{lccc}
\toprule
\textbf{Feature} & \textbf{CD-MAWS~\cite{r1_Anjum,ehsan2025cdmaws}} & \textbf{Peafowl~\cite{r19}} & \textbf{ML-MAWS} \\
\midrule
Sequence feature    & MAWs        & $k$-mers     & MAWs \\
Representation      & Distance\ matrix & Binary\ matrix & Binary\ matrix \\
Tree inference      & NJ          & ML (RAxML)   & ML (IQ-TREE) \\
Strand-aware        & No          & No           & Yes \\
Length selection     & Fixed       & $k$-entropy  & Multi-$\ell$ entropy \\
Model selection     & N/A         & BINGAMMA     & ModelFinder (auto) \\
Ascertainment correction & N/A         & No           & Yes (Mkv) \\
\bottomrule
\end{tabular}
\end{table*}

\subsection{Benchmark Datasets}
\label{sec:datasets}

We tested ML-MAWS on 14 datasets with different genome sizes and varying evolutionary conditions (Table~\ref{tab:datasets}. The AFproject benchmark suite of nine datasets is based on standardized phylogenies of reference to evaluate quantitatively using the standardized AFproject benchmark suite, and the phylogenies are drawn on nine datasets~\cite{r3_Zielezinski}. There are five other viral and mitochondrial whole-genome datasets obtained from the NCBI GenBank \cite{Li2017}.

\begin{table*}[htbp]
\centering
\caption{Benchmark datasets Used in This Study}
\label{tab:datasets}
\begin{tabular}{p{3cm}p{2cm}p{2cm}p{2.3cm}p{2.5cm}}
\toprule
\textbf{Dataset} & \textbf{Type} & \textbf{Species} & \textbf{Avg.\ length} & \textbf{Source} \\
\midrule
Fish mtDNA              & Mitochondrial & 25        & 16.6~kb & AFproject~\cite{r3_Zielezinski} \\
E.\ coli / Shigella    & Bacterial     & 29        & 5.0~Mb  & AFproject \cite{r3_Zielezinski} \\
E.\ coli / Shigella HGT& Bacterial     & 27        & 5.0~Mb  & AFproject \cite{r3_Zielezinski}\\
Yersinia HGT            & Bacterial     & 8         & 4.5~Mb  & AFproject \cite{r3_Zielezinski}\\
Simulated HGT (0--1000) & Simulated     & 33$\times$5 & 1.0~Mb  & AFproject \cite{r3_Zielezinski}\\
\midrule
Coronavirus             & Viral RNA     & 34        & 30~kb   & NCBI GenBank~\cite{Li2017}\\
Ebolavirus              & Viral RNA     & 59        & 19~kb   & NCBI GenBank~\cite{Li2017}\\
Influenza               & Viral RNA     & 38        & 13~kb   & NCBI GenBank~\cite{Li2017}\\
Mammal mtDNA            & Mitochondrial & 41        & 16~kb   & NCBI GenBank~\cite{Li2017}\\
Rhinovirus              & Viral RNA     & 116       & 7.2~kb  & NCBI GenBank~\cite{Li2017}\\
\bottomrule
\end{tabular}
\end{table*}

\subsection{Evaluation Metrics}
\label{sec:evaluation}
In datasets where reference trees are available, we computed the normalized Robinson Foulds (nRF) distance:
\begin{equation}
    \mathrm{nRF}(T, T^*) = \frac{|\sigma(T) \triangle \sigma(T^*)|}{2(m - 3)}
    \label{eq:nrf}
\end{equation}
where $\sigma(T)$ denotes the set of bipartitions induced by the internal edges of tree $T$, $\triangle$ is the symmetric difference, and $2(m-3)$ is the maximum possible RF distance for $m$-taxon unrooted binary trees. The values range from 0 (topologically identical) to 1 (maximally dissimilar). Tree comparisons were performed using DendroPy~\cite{sukumaran2010dendropy}.

In an ablation study, we computed the contribution of the individual components of a pipeline by estimating two variants:
\begin{enumerate}
    \item \textbf{ML-MAWS + IQ-TREE}: full pipeline (strand-aware, multi-length, ModelFinder).
    \item \textbf{ML-MAWS--NoStrand}: ablation removing strand-aware filtering.
\end{enumerate}

\subsection{Implementation}
\label{sec:implementation}

ML-MAWS can be run in C++17, and compiled with GCC~11+. The suffix automaton construction is based on the algorithm of Ehsan et al. \cite{ehsan2025cdmaws}, which is based on the classical construction of Blumer et al. \cite{blumer1985dawg}. Individual species MAW extraction was parallelized over all available CPU cores using OpenMP with dynamic scheduling, showing an almost linear speedup with datasets containing many taxa. Examples of external dependencies are IQ-TREE~v3.1.1 \cite{minh2020iqtree2} for performing ML inference and DendroPy \cite{sukumaran2010dendropy} for comparing trees. All experiments were performed in Google Colab (12 CPU cores, 83GB system RAM).

\section{Results}\label{sec:results}
We tested ML-MAWS on 14 benchmark datasets in terms of normalized Robinson-Foulds distance (nRF), Matching Split Distance (MSD), normalized Quartet Distance (nQD), and bootstrap support. nRF is the fraction of different bipartitions of trees, and MSD provides partial credit to near-correct splits. We measured the quality of datasets without reference trees using bootstrap support. The experiments were conducted on Google Colab (12~cores, 83~GB RAM).

\subsection{Benchmark Datasets}
\label{sec:datasets_results}

The benchmark contains 14 datasets spanning four orders of magnitude in genome size (7~kb to 5~Mb). 9 AFproject datasets~\cite{r3_Zielezinski} provide reference trees: Fish mtDNA (25 species, 16.6~kb), E.\,coli--Shigella (29 species, 5.0~Mb), E.\,coli--Shigella HGT (27 species), Yersinia HGT (8 species, 4.5~Mb), and five simulated datasets with 0--1000 HGT events (33 species, 1.0~Mb each). Five additional datasets from NCBI GenBank lack reference trees: Coronavirus (34 species, 30~kb), Ebolavirus (59 species, 19~kb), Influenza~A (38 species, 13~kb), Mammal mtDNA (41 species, 16~kb), and Rhinovirus (116 species, 7.2~kb).

\subsection{Phylogenetic Accuracy}
\label{sec:accuracy}

\begin{table*}[!t]
\centering
\caption{Phylogenetic accuracy of ML-MAWS variants across AFproject benchmark datasets. nRF and nQD: lower values are better. Bootstrap support: Higher is better. Best values per dataset are shown in \textbf{bold}.}
\label{tab:nrf_results}
\small
\begin{tabular}{ll cccc}
\toprule
\textbf{Dataset} & \textbf{Variant} & \textbf{nRF} $\downarrow$ & \textbf{nQD} $\downarrow$ & \textbf{Avg.\ Bootstrap (\%)} $\uparrow$ & \textbf{\% $\geq$70\%} \\
\midrule
\textit{E.\,coli}--Shigella (29 taxa)
  & IQ-TREE (strand)    & 0.538 & \textbf{0.247} & 83.3 & 84.6 \\
  & NoStrand            & 0.654 & 0.417          & \textbf{88.4} & 80.8 \\
\addlinespace
\textit{E.\,coli}--Shigella HGT (27 taxa)
  & IQ-TREE (strand)    & \textbf{0.583} & \textbf{0.321} & \textbf{92.3} & \textbf{87.5} \\
  & NoStrand            & 0.750 & 0.406          & 79.1 & 70.8 \\
\addlinespace
Fish mtDNA (25 taxa)
  & IQ-TREE (strand)    & 0.409 & 0.185          & 84.3 & 77.3 \\
  & NoStrand            & \textbf{0.182} & \textbf{0.070} & \textbf{96.9} & \textbf{100.0} \\
\addlinespace
Yersinia HGT (8 taxa)
  & IQ-TREE (strand)    & 1.000 & 0.671          & 53.6 & 20.0 \\
  & NoStrand            & 1.000 & \textbf{0.657} & \textbf{63.4} & \textbf{40.0} \\
\addlinespace
Simulated HGT=0 (33 taxa)
  & IQ-TREE (strand)    & \textbf{0.533} & \textbf{0.212} & \textbf{71.1} & \textbf{53.3} \\
  & NoStrand            & 0.733 & 0.374          & 45.7 & 33.3 \\
\addlinespace
Simulated HGT=250 (33 taxa)
  & IQ-TREE (strand)    & 0.700 & \textbf{0.357} & \textbf{62.1} & \textbf{46.7} \\
  & NoStrand            & \textbf{0.633} & 0.358 & 52.9 & 36.7 \\
\addlinespace
Simulated HGT=500 (33 taxa)
  & IQ-TREE (strand)    & 0.733 & \textbf{0.403} & \textbf{46.9} & \textbf{30.0} \\
  & NoStrand            & \textbf{0.667} & 0.435 & 35.5 & 23.3 \\
\addlinespace
Simulated HGT=750 (33 taxa)
  & IQ-TREE (strand)    & 0.733 & \textbf{0.379} & \textbf{45.6} & 26.7 \\
  & NoStrand            & \textbf{0.600} & 0.422 & 44.0 & \textbf{30.0} \\
\addlinespace
Simulated HGT=1000 (33 taxa)
  & IQ-TREE (strand)    & 0.700 & \textbf{0.314} & 33.4 & 20.0 \\
  & NoStrand            & \textbf{0.600} & 0.412 & \textbf{43.0} & \textbf{26.7} \\
\bottomrule
\end{tabular}
\end{table*}

\begin{figure*}[!t]
\centering
\includegraphics[width=\textwidth]{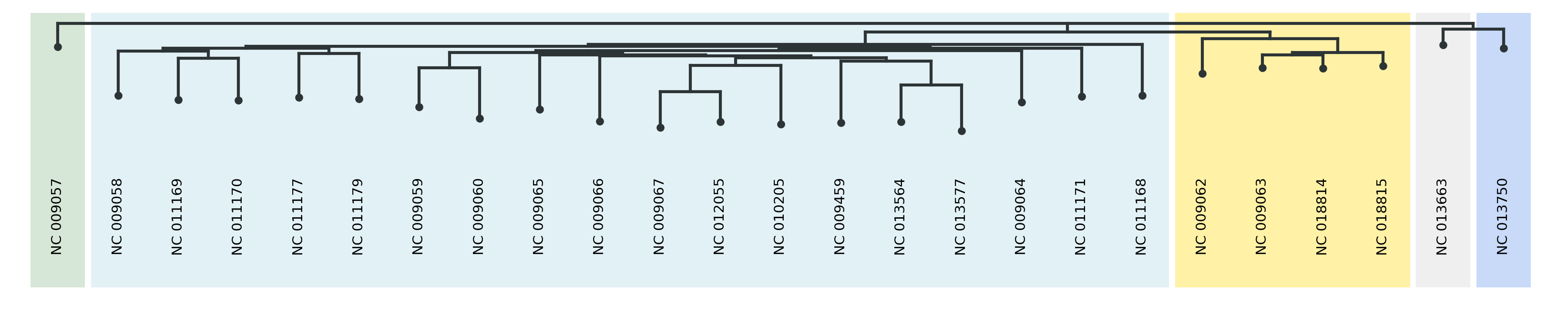}
\caption{ML-MAWS phylogenetic tree for the Fish mtDNA dataset (25 species). Branch labels indicate bootstrap support (\%). The tree recovers major taxonomic groupings with high statistical confidence (nRF\,=\,0.182, avg.\ bootstrap\,=\,96.9\%).}
\label{fig:fish_tree}
\end{figure*}

Table~\ref{tab:nrf_results} summarizes the results across the AFproject datasets. ML-MAWS performed best on the Fish mtDNA dataset, where the NoStrand variant achieved nRF\,=\,0.182, nQD\,=\,0.070, and 96.9\% average bootstrap with 100\% of branches above 70\% (Figure~\ref{fig:fish_tree}). On E.\,coli--Shigella (29 species), ML-MAWS--IQ-TREE obtains nRF\,=\,0.538 with 83.3\% bootstrap; while baselines such as \textit{andi} achieve lower nRF, they cannot provide per-branch confidence. On E.\,coli--Shigella HGT, bootstrap reached 92.3\%---the highest across all bacterial datasets. The Yersinia HGT dataset (8 species, nRF\,=\,1.0) represents the hardest case, although nQD\,=\,0.657 suggests partial quartet-level signal recovery.
ML-MAWS was extremely computationally efficient in the five additional datasets with no reference phylogenies (Coronavirus, Ebolavirus, Influenza A, Mammal mtDNA, Rhinovirus). The entire pipeline took less than 30 s on all five datasets (2 s with Ebolavirus and Rhinovirus, 1 s with Mammal mtDNA) to confirm the practical scalability of the method for small viral and mitochondrial genomes.

\begin{figure*}[!t]
\centering
\includegraphics[width=\textwidth]{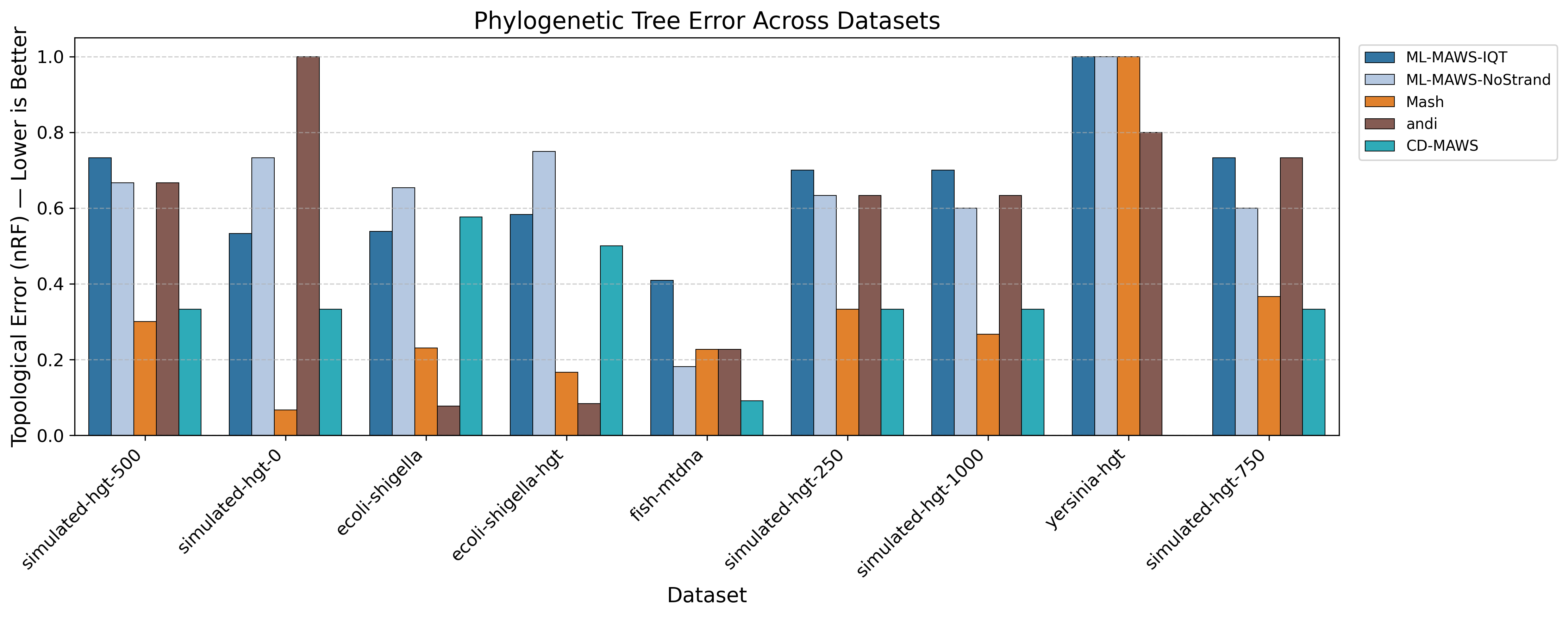}
\caption{Normalized Robinson--Foulds distance (nRF) across benchmark datasets and methods. Lower values indicate better topological accuracy.}
\label{fig:nrf_bar}
\end{figure*}

\begin{figure*}[!t]
\centering
\includegraphics[width=\textwidth]{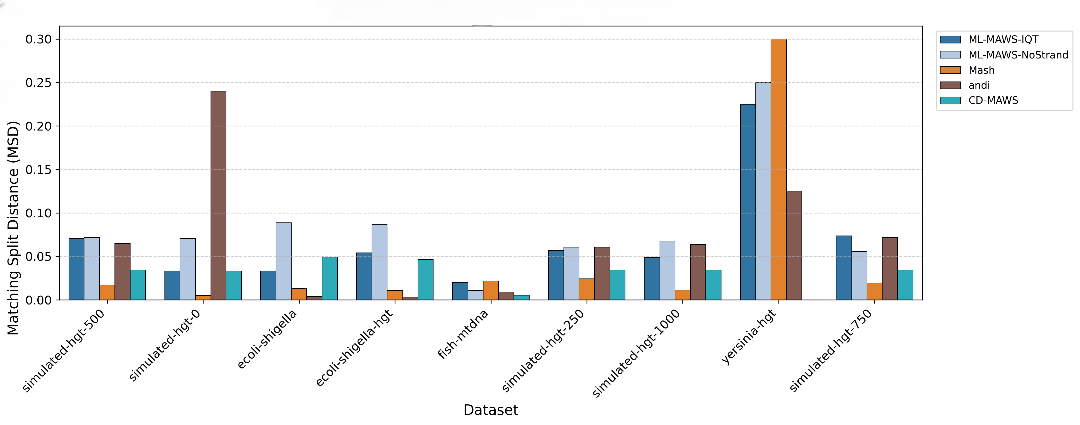}
\caption{Matching Split Distance (MSD) across datasets. MSD provides partial credit for near-correct bipartitions, offering a more nuanced view of topological accuracy than the binary nRF metric. Lower values are better.}
\label{fig:msd_bar}
\end{figure*}

\subsection{Why ML-MAWS Exhibits Higher nRF and MSD Than Existing Baselines}
\label{sec:why_error}
ML-MAWS achieves higher nRF and MSD values than the compared baselines across the benchmark datasets (Figures~\ref{fig:nrf_bar} and~\ref{fig:msd_bar}). This is a direct consequence of the representational trade-off inherent in the design of the method. The baselines use continuous valued features, such as Minhash sketches, spaced-word frequencies, or cosine distances over composition vectors, which provide fine-grained quantitative differences between sequences and pass them to NJ to construct a tree. In ML-MAWS, instead, each Minimal Absent Word is represented by a binary character, either present or absent, without frequency and compositional information. 

Binary encoding is also inherently coarser than continuous-valued representations; therefore, it is harder to recover an exact bipartition, especially for closely related species in which few MAWs differ. The nRF metric compounds this difference because it is a binary measure that gives full credit for bipartition if it is correct, but no credit for them if it is wrong, even if it is a near-correct split. The MSD results showed that most of the incorrect bipartitions of ML-MAWS were one or two taxa away from the reference, meaning that the topologies of these bipartitions were just rearranging a few taxa, not grossly misplacing the entire phylogenetic tree. Here, the Yersinia HGT dataset was too small for the CD-MAWS cosine distance measure to obtain a reliable tree.
Importantly, ML-MAWS also recovers better topologies than the distance-based CD-MAWS approach (e.g., nRF 0.538 vs.\ 0.577 on E.\,coli--Shigella) in the same MAW feature space, supporting that binary character encoding with maximum likelihood inference outperforms pairwise distance aggregation with the NJ method. 

Moreover, nRF and MSD alone are insufficient for evaluating a phylogenetic method. ML-MAWS is the only method in this comparison that provides per-branch bootstrap support, operates under a principled statistical model with ascertainment bias correction and automatic model selection, and produces well-calibrated confidence values that correctly reflect increasing phylogenetic uncertainty capabilities that none of the compared baselines can offer. In clinical genomics and epidemiological surveillance, trees without confidence measures are scientifically incomplete, making the statistical framework of ML-MAWS essential despite its topological accuracy trade-off.

\subsection{Ablation Study: Strand-Aware Filtering}
\label{sec:ablation}

\begin{figure*}[!t]
\centering
\includegraphics[width=\textwidth]{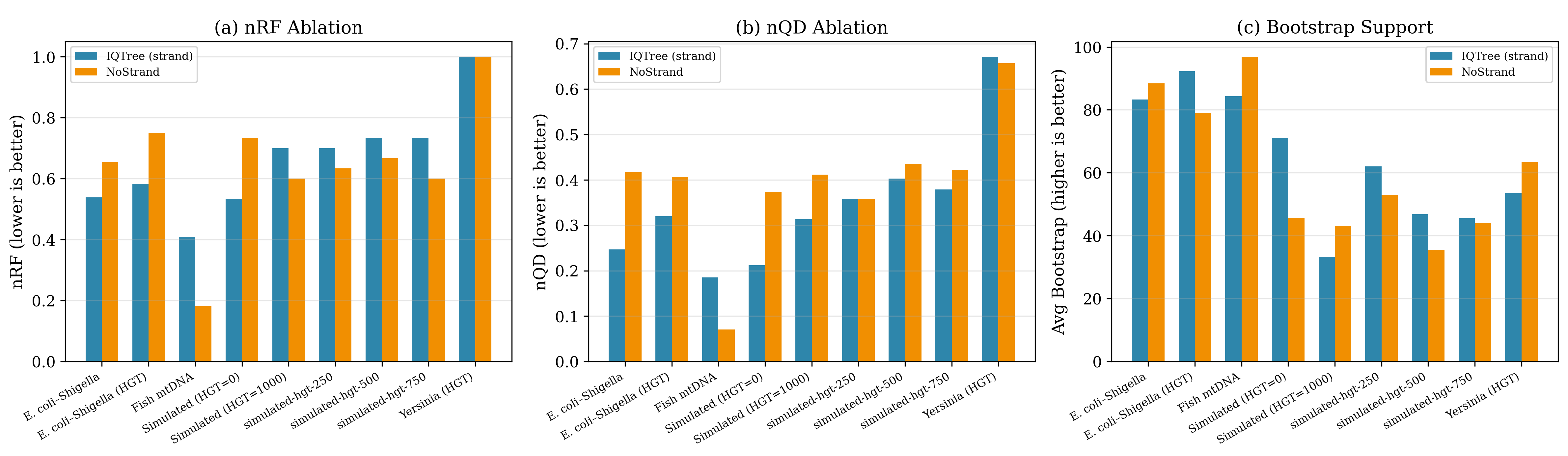}
\caption{Ablation study comparing strand-aware (IQ-TREE) and no-strand variants across datasets: (a)~nRF, (b)~nQD, and (c)~bootstrap support.}
\label{fig:ablation}
\end{figure*}

Figure~\ref{fig:ablation} compares the strand-aware and NoStrand variants. Strand filtering improves nQD by 17.8\% on average, with the largest gain on E.\,coli--Shigella HGT (nQD: 0.406$\to$0.321, 20.9\% improvement). MSD supports this when nQD improves, and MSD also decreases. On Fish mtDNA, disabling strand filtering improved nRF from 0.409 to 0.182, as mitochondrial genomes lack the replication-associated strand asymmetry of bacterial chromosomes.

\subsection{Impact of Horizontal Gene Transfer}
\label{sec:hgt}

\begin{figure}[htbp]
\centering
\includegraphics[width=0.5\textwidth]{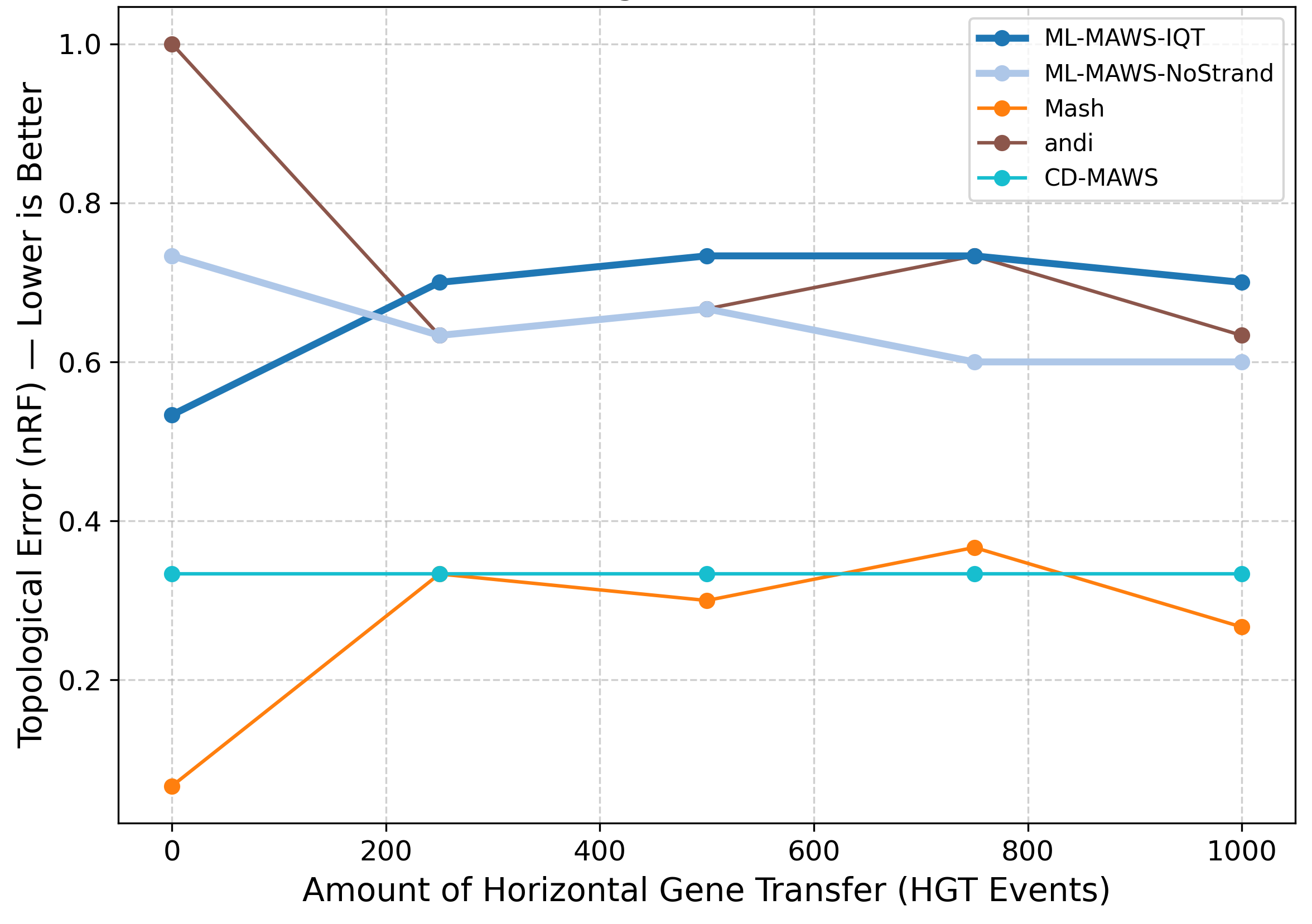}
\caption{Effect of increasing horizontal gene transfer (HGT) on nRF across five simulated datasets (0, 250, 500, 750, and 1000 HGT events). ML-MAWS variants are shown alongside baseline methods.}
\label{fig:hgt}
\end{figure}

Figure~\ref{fig:hgt} shows accuracy degradation with increasing HGT. For the strand-aware variant, nRF rises from 0.533 (HGT=0) to 0.733 (HGT=500), stabilizing at 0.700 (HGT=1000). Bootstrap decreases monotonically: 71.1\%$\to$62.1\%$\to$46.9\%$\to$45.6\%$\to$33.4\%, correctly reflecting the increasing topological uncertainty. Distance-based methods cannot convey these signals.

\subsection{Bootstrap Support as a Distinguishing Feature}
\label{sec:bootstrap}

ML-MAWS uniquely provides per-branch statistical confidence; 83.3\% on E.\,coli--Shigella, 92.3\% on E.\,coli HGT, 84.3\% on Fish mtDNA, and 96.9\% (NoStrand) with 100\% of branches above 70\%. No distance-based baseline (Mash, Skmer, andi, CD-MAWS, FFP, Co-phylog, kSNP) could produce bootstrap values. In clinical and epidemiological contexts, quantifying confidence is essential for sound decision-making.

\subsection{Computational Performance}
\label{sec:performance}

\begin{figure*}[htbp]
\centering
\includegraphics[width=\textwidth]{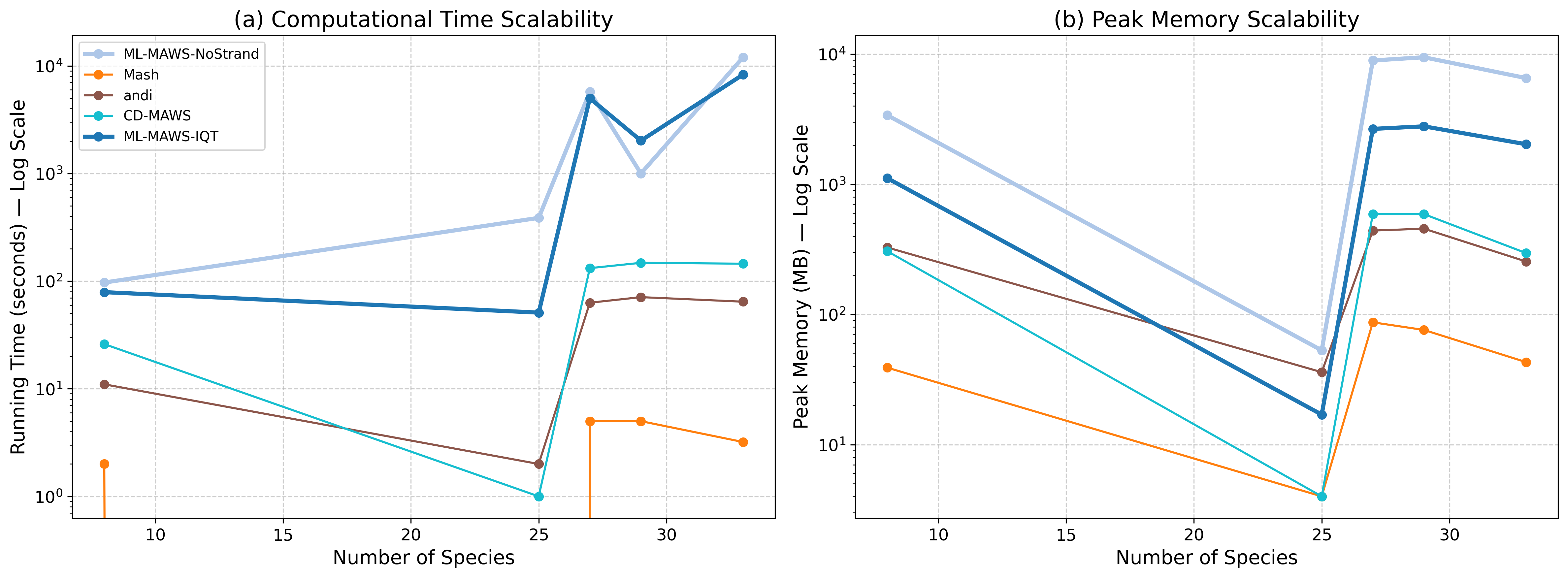}
\caption{Log-scale comparison of (a)~running time and (b)~peak memory across all methods. ML-MAWS trades computational resources for statistical rigour through maximum likelihood inference.}
\label{fig:scalability}
\end{figure*}

For Fish mtDNA, the full pipeline was completed in 51\,s with 17\,MB memory (Table~\ref{tab:timing}). On E.\,coli--Shigella, ML inference dominated (1,723 of 2,020\,s, 85\%). The suffix automaton-based MAW extraction is theoretically optimal and highly efficient: for $m$ sequences of average length $n$ over an alphabet of size $\sigma$, feature extraction requires $\mathcal{O}(m \cdot n \cdot \sigma)$ total time and $\mathcal{O}(n \cdot \sigma)$ peak memory per sequence. Empirically, this takes just 238\,s (12\% of total time) for the 5.0\,Mb dataset. Subsequent construction of the binary character matrix requires $\mathcal{O}(m \cdot |U|)$ space, where $|U|$ is the number of unique informative MAWs. Distance-based methods (Mash: $<$5\,s, andi: 72\,s) are faster but lack bootstrap support (Figure~\ref{fig:scalability}). ML-MAWS trades the $\mathcal{O}(n \log n)$ speed of alignment-free distances for the theoretically superior but NP-hard maximum likelihood tree inference, offering statistically supported trees.

\begin{table}[!t]
\centering
\caption{Per-step time breakdown (seconds) and peak memory (MB) for ML-MAWS--IQ-TREE on representative datasets.}
\label{tab:timing}
\small
\begin{tabular}{p{1.2cm}p{0.4cm}p{0.5cm}rrrr}
\toprule
\textbf{Dataset} & \textbf{Read} & \textbf{MAW} & \textbf{Entropy} & \textbf{Matrix} & \textbf{ML} & \textbf{Mem.} \\
\midrule
Fish mtDNA      & 8   & $<$1  & $<$1  & $<$1  & 42    & 17 \\
E.\,coli--Shig. & 16  & 238   & 18    & 25    & 1,723  & 2,788 \\
\bottomrule
\end{tabular}
\end{table}

\subsection{Entropy-Based Length Selection}
\label{sec:entropy_results}

The entropy criterion automatically selects optimal MAW lengths: $\ell^*\!=\!11$ with aggregation over $\{11,12,13\}$ for E.\,coli--Shigella (range $[5,14]$), and $\ell^*\!=\!7$ over $\{7,8,9\}$ for Fish mtDNA (range $[3,10]$). This eliminates manual parameter tuning across genome sizes that span four orders of magnitude.

\section{Discussion and Future Work}
\label{sec:discussion}
ML-MAWS combines different paradigms for MAW-based phylogenetics rather than compressing MAW information into pairwise distances. It constructs a binary character matrix suitable to ML inference under the Lewis Mkv model with ascertainment bias correction. This enables model selection, bootstrapping, and proper statistical estimation features unavailable in all previous MAW-based approaches.

We identify three scenarios in which this framework is most valuable. First, on mitochondrial/viral genomes, it achieves excellent accuracy with high confidence (Fish mtDNA: nRF\,=\,0.182, 96.9\% bootstrap, $<$1\,min). Second, for HGT-affected genomes, it provides calibrated confidence (E.\,coli HGT: 92.3\% bootstrap, 87.5\% nodes $\geq$70\%). Third, strand-aware filtering improves nQD and MSD in bacterial genomes by removing reverse-complement artifacts.
We showed that distance-based methods achieved lower nRF on closely related bacteria (\textit{andi}: 0.077, Mash: 0.231 vs.\ ML-MAWS: 0.538), as binary MAW encoding lacks the quantitative resolution of continuous-valued features for species differing by few substitutions. ML-MAWS is also slower ($\sim$33\,min vs.\ $<$5\,s for Mash) because of the NP-hard ML search. Branch lengths represent MAW state transitions, not nucleotide substitutions, making the WRF inapplicable. 

Despite the topological accuracy trade-off, ML-MAWS provides capabilities that no existing alignment-free method can offer: (1)~it is the only MAW-based method providing per-branch bootstrap support essential in clinical genomics where trees without confidence measures are scientifically incomplete; (2)~MSD reveals that many incorrect bipartitions are near-correct splits differing by 1--2 taxa; (3)~the Lewis Mkv framework with ModelFinder provides a principled statistical foundation absent from heuristic distances; (4)~bootstrap values are well-calibrated, decreasing monotonically with HGT (71.1\%$\to$33.4\%); (5)~ML-MAWS outperforms CD-MAWS (0.538 vs.\ 0.577), demonstrating that character matrix + ML inference improves upon distance + NJ within the MAW paradigm.

Compared to Peafowl~\cite{r19}, the closest existing method, ML-MAWS employs linguistically motivated MAWs rather than generic $k$-mers, ascertainment bias correction, and strand-aware intersection filtering. Overall, baselines can achieve lower nRF on certain datasets, but they cannot give the statistical rigor and quantification of confidence that are needed to perform responsible phylogenetic inference in clinical and epidemiological applications. The future directions consist of weighted MAW character encoding to better resolve similar species and heterogeneous partition models in MAW length classes, SA construction on a GPU, and longer features (shortest unique substrings, rare $k$-mers).

Future work will explore heuristic tree rearrangement methods (e.g., stochastic nearest-neighbor interchange, subtree pruning and regrafting) and approximation methods to overcome the NP-hardness of the ML tree search for larger taxon sets, where the tree search will be significantly faster and only have a bounded topological error, making ML-MAWS suitable for large datasets of hundreds or thousands of species. In addition, we will investigate the Pairwise Comparability Graph (PCG) representation where each node is a biological sequence and each edge represents a pair of sequences that can be distinguished by at least one MAW from a given set of MAW lengths. The study of the structural properties of this graph will help to formally characterize how well a set of chosen MAW lengths separates the input sequences, thereby helping to choose informative lengths of the MAW and increasing the discriminative power of the binary character matrix. 

\section{Conclusion}\label{sec:conclusion}

We introduced ML-MAWS, the first MAW-based phylogenetic method that provides per-branch bootstrap support within the maximum likelihood framework. ML-MAWS can generate phylogenetic trees with statistical confidence on each branch of the tree by encoding each individual MAW as a binary character and estimating the tree using the Lewis Mkv model with ascertainment bias correction.

A comparison with 14 benchmark datasets showed that ML-MAWS achieves strong topological accuracy, especially on mitochondrial-scale genomes, where it achieves nRF\,=\,0.182 and 96.9\% average bootstrap support. The strand-aware intersection filtering step incurred an average 17.8\% improvement in quartet-level accuracy on bacterial genomes, and the entropy-based multi-length selection removed the need to manually tune the parameters, enabling the method to automatically scale to genome sizes of four orders of magnitude. Although distance-based methods achieve better topological accuracy on closely related bacterial genomes, they cannot provide the per-branch statistical confidence that ML-MAWS uniquely offers, a capability essential for responsible phylogenetic inference in epidemiology, clinical genomics, and regulatory phylogenomics.

\section*{Research Ethics Declaration}
\subsection*{Originality and Plagiarism Declaration}

I on behalf of my group hereby declare that this manuscript is our original work and has not been plagiarized. All sources of information have been properly acknowledged and cited. This work has not been submitted, in whole or in part, to any other journal, conference, or academic institution for degree or publication.

\subsection*{Authorship Declaration}

All individuals listed as authors have made substantial contributions to the conception, design, data collection, analysis, or interpretation of the study and approve the final version of the manuscript. No eligible contributor has been omitted, and no ineligible individual has been included.

\subsection*{Data Integrity and Accuracy}

We affirm that the data presented in this manuscript are accurate, complete, and honestly reported. No data have been fabricated, falsified, or inappropriately manipulated.

\subsection*{Data Availability and Confidentiality}

All data used in this study are publicly available benchmark datasets from the AFproject~\cite{r3_Zielezinski} and NCBI GenBank~\cite{Li2017}. No confidential or personally identifiable information was used. The source code and all scripts required to reproduce the experiments are available under the Apache licence at: \url{https://github.com/PapriSaha/ML-MAWS}.

\subsection*{Artificial Intelligence (AI) Use Declaration}

Artificial intelligence tools were used only for language editing and formatting. No AI tool was used to generate original research data, analysis, or conclusions. The authors take full responsibility for the content of this manuscript.

\subsection*{Conflict of Interest Declaration}

We declare that there are no financial, personal, or academic conflicts of interest that could have influenced the work reported in this manuscript.

\subsection*{Funding Declaration}

This research received no external funding.

\subsection*{Student Declaration (Supervisor-Approved Work)}

This work was carried out under the supervision of Dr.\ Md.\ Manzurul Hasan, Associate Professor, CS, American International University-Bangladesh (AIUB), Bangladesh. Any assistance received has been acknowledged. The responsibility for errors or omissions remains entirely with the authors.

\subsection*{Final Ethics Declaration Statement}

We hereby confirm that this research complies with all applicable ethical standards, institutional guidelines, and academic integrity requirements. We accept full responsibility for the content of this manuscript.

\section*{Acknowledgment}
We would like to thank Prof.\ Dr.\ Md.\ Shamsuzzoha Bayzid, Dept.\ of CSE, BUET for some initial discussion, guidance, and comments.

\ifCLASSOPTIONcaptionsoff
  \newpage
\fi

\begin{IEEEbiography}[{\includegraphics[width=1in,height=1.25in,clip,keepaspectratio]{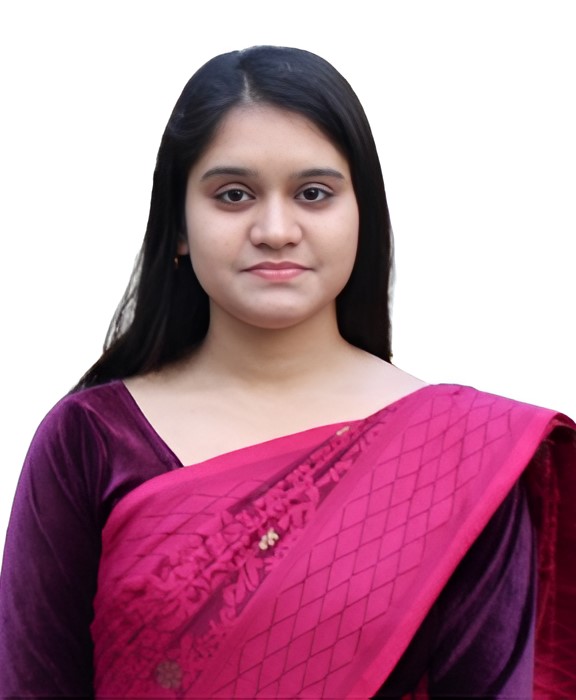}}]{Papri Saha}
is a graduate student currently pursuing a Master of Science in Computer Science (Intelligent Systems) at the American International University-Bangladesh (AIUB), Dhaka in the Department of Computer Science. She completed her Bachelor of Science in Computer Science and Engineering at AIUB. Her research interests include advanced fields such as machine learning, artificial intelligence, reinforcement learning, federated learning, graph neural networks, computational biology, and bioinformatics. She has focused on developing innovative algorithms that advance the theoretical foundations and practical implementation of computational intelligence, distributed systems, and bioinformatics algorithms.
\end{IEEEbiography}

\vskip -2\baselineskip plus -1fil

\begin{IEEEbiography}[{\includegraphics[width=1in,height=1.25in,clip,keepaspectratio]{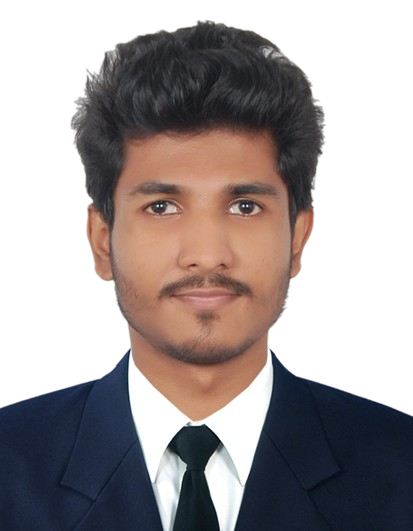}}]{Sudipta Kumar Das}
received his B.Sc. degree in Computer Science and Engineering from American International University-Bangladesh (AIUB) and his Professional Master of Science in Computer Science (PMSCS) from Jahangirnagar University (JU). He is currently pursuing his M.Sc. degree in Computer Science at American International University-Bangladesh (AIUB).  
He is currently serving as an Academic Instructor in the Department of Computer Science at American International University-Bangladesh (AIUB). His professional and academic background includes industry experience as a software engineering intern with EnKaizen and Itransition, alongside teaching undergraduate courses in Introduction to Programming and Data Structures. His research and development interests include bioinformatics, design and analysis of algorithms, practical data structures, systems programming, and tool development using emerging languages such as Rust.
\end{IEEEbiography}

\vskip -2\baselineskip plus -1fil

\begin{IEEEbiography}[{\includegraphics[width=1in,height=1.25in,clip,keepaspectratio]{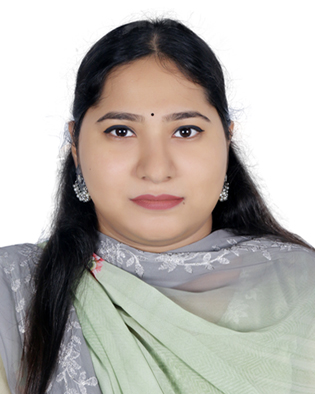}}]{Anonnya Sarkar} received her B.Sc. degree in Computer Science and Engineering and is currently pursuing her M.Sc. degree in Computer Science and Engineering from American International University-Bangladesh (AIUB). She is currently working as an IT Executive at Intellier Limited where she is serving a Japanese power plant company (JERA). In addition to her professional responsibilities, she is actively involved in academic research. Her current research focuses on machine learning, bioinformatics and cyber security.
\end{IEEEbiography}

\vskip -2\baselineskip plus -1fil

\begin{IEEEbiography}[{\includegraphics[width=1in,height=1.25in,clip,keepaspectratio]{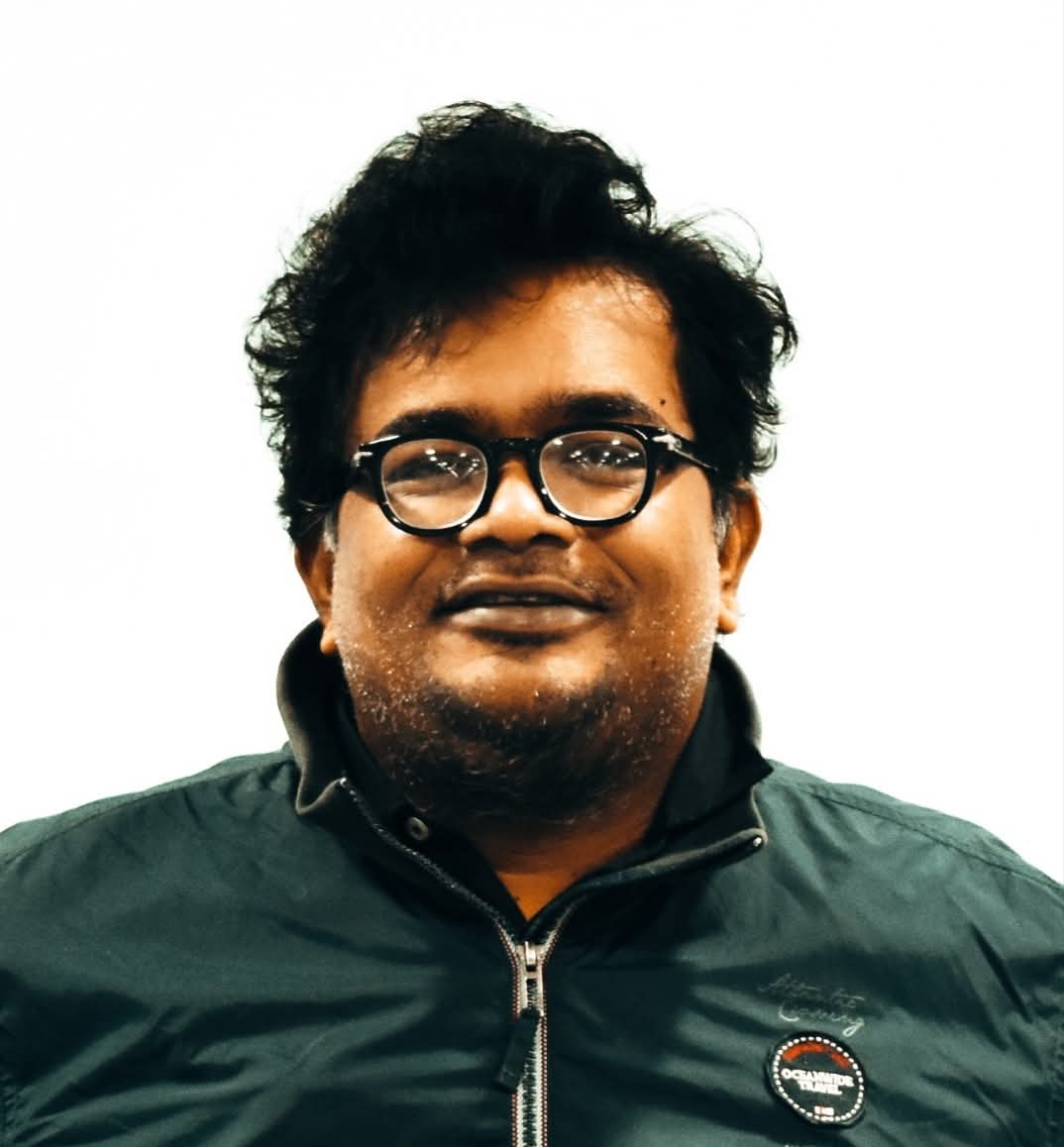}}]{Md. Manzurul Hasan} received his Ph.D. from the Department of Computer Science and Engineering (CSE), BUET, Bangladesh. Dr. Hasan obtained his M. Sc. Engg. in CSE and B. Sc. Engg. in Computer Science and Engineering (CSE) and BUET, Bangladesh and CUET, Bangladesh respectively. He is currently serving at the Department of Computer Science as an Associate Professor at American International University-Bangladesh (AIUB). He served at different reputed professional organizations and different teaching positions with a service length of about 20 years. He is an active member of different professional bodies. Depending on the research activity at his young age, he has been awarded JSPS HOPE (Nobel Laureates) Fellow, Yokohama, Japan, 2025. Dr. Hasan has many prestigious publications in different journals and different conference proceedings published by different prestigious publishers. His research interests include Algorithms fundamentals, NP-hard problems, Approximation algorithms, Online algorithms, Distributed algorithms, String algorithms, Heuristics \& Metaheuristics, Graph Theory, Computer networks' protocols and topologies, Database and Data management systems, Information visualization, Management Information Systems \& Information System Management, Security in Computing etc.
\end{IEEEbiography}




\end{document}